\documentclass[11pt,letterpaper]{JHEP3}
\usepackage{graphics}
\usepackage{epsfig}
\usepackage{subfigure}
\usepackage{mathrsfs}
\usepackage{amsmath}
\usepackage{amsfonts}
\usepackage{amssymb}
\usepackage{graphicx}
\usepackage{longtable}
\usepackage{rotating}
\usepackage{multirow}
\title{ Van der Waals phase transition in the framework of holography}
\author{Xiao-Xiong Zeng$^{1,2}$, Li-Fang Li$^{3}$\\
$^{1}$ School of Material Science and Engineering, Chongqing Jiaotong University, Chongqing,\\
            ~ 400074, China\\
$^{2}$ Institute of Theoretical Physics,
       Chinese Academy of Sciences,
       Beijing 100190, China\\
$^{3}$ State Key Laboratory of Space Weather, Center for Space Science and Applied Research,\\
       Chinese Academy of Sciences, Beijing 100190, China\\
 \email{xxzeng@itp.ac.cn;  lilf@itp.ac.cn}
}
 \abstract{Phase structure of the    quintessence  Reissner-Nordstr\"{o}m-AdS  black hole  is probed by
the nonlocal observables such as   holographic  entanglement entropy and two point correlation function. Our result shows that,
 as the case of the  thermal entropy, both the   observables  exhibit the Van der Waals-like phase transition.
To reinforce this conclusion, we  further check  the equal area law for the first order phase transition and critical
exponent of the heat capacity for the second order phase transition.
  We also discuss the effect of the state parameter on the phase structure of the nonlocal observables.

}

\begin{document}
\bibliographystyle{ieeetr}
\bibliography{reference.bib}

\begin{thebibliography}{100}



\bibitem{Hawking4} S. W. Hawking and D. N. Page, \emph{Thermodynamics Of Black Holes In Anti-De Sitter
Space}, Commun. Math. Phys. 87, 577 (1983)


\bibitem{Witten1} E. Witten, \emph{Anti-de Sitter space, thermal phase transition, and confinement in gauge
theories}, Adv. Theor. Math. Phys. 2, 505 (1998) [arXiv:hep-th/9803131]

\bibitem{Chamblin} A. Chamblin, R. Emparan, C. V. Johnson and R. C. Myers, \emph{Charged AdS black holes
and catastrophic holography}, Phys. Rev. D 60, 064018 (1999) [hep-th/9902170]

\bibitem{Kubiznak} D. Kubiznak and R. B. Mann,\emph{ P-V criticality of charged AdS black holes}, JHEP 1207, 033 (2012) [arXiv:1205.0559 [hep-th]]


\bibitem {Xu}  J. Xu, L. M. Cao, Y. P. Hu, \emph{P-V criticality in the extended phase space of black holes in massive gravity}, Phys. Rev. D 91, 124033 (2015) [arXiv:1506.03578 [gr-qc]]

\bibitem {Caipv} R. G. Cai, L. M. Cao, L. Li, R. Q. Yang, \emph{P-V criticality in the extended phase space of
Gauss-Bonnet black holes in AdS space}, JHEP 09, 005 (2013) [arXiv:1306.6233 [gr-qc]]


\bibitem {Hendi}  S. H. Hendi, A. Sheykhi, S. Panahiyan, B. Eslam Panah, \emph{Phase transition and thermodynamic geometry of Einstein-Maxwell-dilaton black holes}, Phys. Rev. D 92, 064028 (2015) [arXiv:1509.08593 [hep-th]]


\bibitem {Hennigar} R. A. Hennigar, W. G. Brenna, R. B. Mann, \emph{P-V Criticality in Quasitopological Gravity}, JHEP 1507, 077 (2015) [arXiv:1505.05517 [hep-th]]

\bibitem {Wei1} S. W. Wei, Y. X. Liu, \emph{Clapeyron equations and fitting formula of the coexistence curve in the extended phase space of charged AdS black holes},  Phys. Rev. D 91, 044018 (2015) [arXiv:1411.5749 [hep-th]]

\bibitem {Mo} J. X. Mo, W. B. Liu, \emph{P-V Criticality of Topological Black Holes in Lovelock-Born-Infeld Gravity}, Eur. Phys. J. C 74, 2836 (2014) [arXiv:1401.0785 [hep-th]]

\bibitem {Spallucci} E. Spallucci and A. Smailagic, Maxwell's equal area law for charged Anti-deSitter black
holes, Phys. Lett. B 723, 436 (2013) [arXiv:1305.3379 [hep-th]]




\bibitem {Johnson}  C. V. Johnson, \emph{ Large N Phase Transitions, Finite Volume, and Entanglement Entropy}, JHEP 1403, 047 (2014) [arXiv:1306.4955 [hep-th]]



\bibitem {Caceres}  E. Caceres, P. H. Nguyen, J. F. Pedraza, \emph{Holographic entanglement entropy and the extended phase structure of STU black holes}, JHEP 1509, 184, (2015) [arXiv:1507.06069 [hep-th]]


\bibitem {Nguyen}  P. H. Nguyen, \emph{An equal area law for the van der Waals transition of holographic entanglement entropy}, JHEP 12, 139 (2015) [arXiv:1508.01955 [hep-th]]


\bibitem {zeng2016} X. X. Zeng, H. Zhang, L. F. Li, \emph{Phase transition of holographic entanglement entropy in massive gravity} [arXiv:1511.00383 [gr-qc]]


\bibitem {Dey} A. Dey, S. Mahapatra, T. Sarkar,\emph{ Thermodynamics and Entanglement Entropy with Weyl Corrections},  [arXiv:1512.07117 [hep-th]]


 \bibitem{Balasubramanian1} V. Balasubramanian {\it et al.},\emph{ Thermalization of Strongly Coupled Field
Theories}, Phys. Rev. Lett. 106, 191601 (2011) [arXiv:1012.4753 [hep-th]]

\bibitem{Balasubramanian2} V. Balasubramanian {\it et al.}, \emph{Holographic Thermalization}, Phys. Rev. D 84, 026010 (2011) [arXiv:1103.2683 [hep-th]]

\bibitem{GS}D. Galante and M. Schvellinger, \emph{Thermalization with a chemical potential from AdS
spaces}, JHEP 1207, 096 (2012)  [arXiv:1205.1548 [hep-th]]

\bibitem{CK}E. Caceres and A. Kundu,   \emph{ Holographic Thermalization with Chemical Potential}, JHEP 1209, 055 (2012) [arXiv:1205.2354 [hep-th]]


\bibitem{Zeng2013}  X. X. Zeng, B. W. Liu, \emph{Holographic thermalization in Gauss-Bonnet gravity},  Phys. Lett. B 726,   481 (2013) [arXiv:1305.4841 [hep-th]]

\bibitem{Zeng2014}  X. X. Zeng, X. M. Liu, B. W. Liu, \emph{Holographic thermalization with a chemical potential in Gauss-Bonnet gravity}, JHEP 03, 031 (2014)  [arXiv:1311.0718 [hep-th]]




\bibitem{Superconductors1} T. Albash, C. V. Johnson, \emph{Holographic Studies of Entanglement Entropy in Superconductors}, JHEP 1205, 079 (2012), [arXiv:1202.2605 [hep-th]]

\bibitem{Superconductors2}
R. G. Cai, S. He, L. Li, Y. L. Zhang,\emph{ Holographic Entanglement Entropy in Insulator/Superconductor Transition}, JHEP 1207, 088 (2012), [arXiv:1203.6620 [hep-th]]

\bibitem{Superconductors3}
R. G. Cai, L. Li, L. F. Li, R. K. Su, \emph{Entanglement Entropy in Holographic P-Wave Superconductor/Insulator Model}, JHEP 1306, 063 (2013), [arXiv:1303.4828 [hep-th]]

\bibitem{Superconductors4}
L. F. Li, R. G. Cai, L. Li, C. Shen,\emph{ Entanglement entropy in a holographic p-wave superconductor model}, Nucl. Phys. B 894, 15-28 (2015), [arXiv:1310.6239 [hep-th]]

\bibitem{Superconductors5}
R. G. Cai, S. He, L. Li, Y. L. Zhang,\emph{\emph{ Holographic Entanglement Entropy in Insulator/Superconductor Transition}}, JHEP 1207, 088 (2012),  [arXiv:1203.6620 [hep-th]]


\bibitem{Superconductors7}
X. Bai, B. H. Lee, L. Li, J. R. Sun, H. Q. Zhang, \emph{Time Evolution of Entanglement Entropy in Quenched Holographic Superconductors}, JHEP 040, 66 (2015), [arXiv:1412.5500 [hep-th]]

\bibitem{Superconductors8}
R. G. Cai, L. Li, L. F. Li, R. Q. Yang, \emph{Introduction to Holographic Superconductor Models}, Sci China-Phys Mech Astron, 58, 060401 (2015),  [arXiv:1502.00437 [hep-th]]


\bibitem{Ling} Y. Ling, P. Liu, C. Niu, J. P. Wu, Z. Y. Xian, \emph{Holographic Entanglement Entropy Close to Quantum Phase
Transitions}, [arXiv:1502.03661 [hep-th]]






\bibitem{Engelhardt} N. Engelhardt, T. Hertog, G. T. Horowitz, \emph{Holographic Signatures of Cosmological Singularities}, Phys. Rev. Lett. 113, 121602 (2014) [arXiv:1404.2309  [hep-th]]




\bibitem{Engelhardt1} N. Engelhardt, T. Hertog, G. T. Horowitz, Further Holographic Investigations of Big Bang Singularities, JHEP 1507,  044 (2015)  [arXiv:1503.08838 [hep-th]]

\bibitem{Sb1} S. Chen, B. Wang, R. Su, \emph{Hawking radiation in a $d$-dimensional static spherically-symmetric black Hole surrounded by quintessence}, Phys.Rev. D 77, 124011 (2008)

\bibitem{Sb2} S. Chen, J. Jing,\emph{ Quasinormal modes of a black hole surrounded by quintessence}, Class. Quant. Grav.  22,
 4651 (2005)

\bibitem{Chen:2012mva}  S.~Chen, Q.~Pan and J.~Jing,\emph{ Holographic superconductors in quintessence AdS black hole spacetime},
Class. Quant. Grav. 30,  145001 (2013) [arXiv:1206.2069 [gr-qc]]



\bibitem{Zeng20151} X. X. Zeng, D. Y. Chen, L. F. Li,\emph{ Holographic thermalization and gravitational collapse in the spacetime dominated by quintessence dark energy}, Phys. Rev. D 91, 046005 (2015)   [arXiv:1408.6632 [hep-th]]

\bibitem{Libh} G. Q. Li,\emph{ Effects of dark energy on P-V criticality of charged AdS black holes}, Physics Letters B 735, 256-260 (2014) [arXiv:1407.0011 [gr-qc]]



\bibitem{Kiselev} V. V. Kiselev, Quintessence and black holes, \emph{Class. Quan. Grav. } 20 (2003) 1187




\bibitem{Ryu} S. Ryu and T. Takayanagi, \emph{Holographic derivation of entanglement entropy from
AdS/CFT}, Phys. Rev. Lett. 96, 181602 (2006)  [hep-th/0603001]

\bibitem{Ryu1}   S. Ryu, T. Takayanagi,\emph{ Aspects of Holographic Entanglement Entropy}, JHEP
0608, 045 (2006)  [hep-th/0605073]



\bibitem{Balasubramanian61}    V. Balasubramanian and S. F. Ross,\emph{ Holographic particle detection}, Phys. Rev. D 61,
044007 (2000)  [arXiv:hep-th/9906226]


\end{thebibliography}

\section{Introduction}

Investigation on the thermodynamic phase transition of a black hole is always a hot topic in black hole physics. On  one hand, It is helpful for us to understand the nature of some quantities of  black holes such as entropy. On the other hand, it also may shed light on the  understanding of the relation between gravity and thermodynamics. Until now there are many works to study the phase transition of
black holes. The AdS space time is the most popular background for it exhibits more abundant phase structures comparing with its counterparts. The particular phase transition in AdS space time is the  Hawking-Page phase transition between the AdS black hole and thermal gas \cite{Hawking4}, which is interpreted as the confinement/deconfinement
phase transition in the dual gauge field theory \cite{Witten1}. Another interesting phenomenon for  a charged  AdS black hole
is that it exhibits the Van der Waals-like phase transition in the $T-S$ plane \cite{Chamblin}. Namely the black holes endowed with different
charges have different phase structures.
As the charge increases from small to large, the hole
will undergo first order phase transition and second order phase transition successively  before it reaches to a stable phase. Recently in the extended phase space, where the negative cosmological constant is treated as the  pressure
while its conjugate acts as the thermodynamical volume, the Van der Waals-like phase transition is reconstructed in the $P-V$
plane \cite{Kubiznak,Xu,Caipv,Hendi,Hennigar,Wei1,Mo}. It was stressed  that there is a duality  between the Van der Waals-like phase transition in  $P-V$
plane and  $T-S$ plane similar to the T-duality of string theory \cite{Spallucci}.

In this paper, we intend to study the Van der Waals-like phase transition in the framework of holography. Our work is based on the previous study in \cite{Johnson} where the phase structure of entanglement entropy is studied in a fixed charge ensemble and chemical potential ensemble. They found that the phase structure of entanglement entropy was similar to that of the thermal entropy for a charged black hole. They also studied the critical behavior of the heat capacity  in the neighborhood of the critical point and found that the critical exponent was the same as the one of the thermal entropy for the second order phase transition of holographic entanglement entropy.
 In light of these interesting results,   this work was extended to the extended phase space later, where the cosmological
constant was treated as a thermodynamical variable. It was found that the entanglement entropy has the similar phase structure as that of the black hole entropy too \cite{Caceres}. In \cite{Nguyen}, Nguyen investigated exclusively the equal area law of holographic entanglement entropy and found that, as the case of thermal entropy, the equal area law  holds  for the entanglement entropy regardless of the size of the entangling region. Very recently \cite{zeng2016} investigated entanglement entropy for a  quantum system with infinite volume, and \cite{Dey}  investigated entanglement entropy in the bulk with Weyl correction,
both of their result showed that  there ia a  Van der Waals-like phase transition in the entanglement entropy-temperature plane.

Note that in all the works mentioned above, the authors considered only the phase structure of entanglement entropy in the field theory. In this paper, we will further study the  phase structure of two point correlation function besides that of the entanglement entropy. The two point correlation function is also a nonlocal observable and to some extent it has the similar properties as the entanglement entropy. For example, both of them can probe the non-equilibrium thermalization behavior \cite{
Balasubramanian1,Balasubramanian2,GS,CK, Zeng2013, Zeng2014}, superconductor phase transition \cite{
Superconductors1, Superconductors2, Superconductors3, Superconductors4, Superconductors5, Superconductors7, Superconductors8,Ling}, and cosmological singularity \cite{Engelhardt,Engelhardt1}.  In this paper, we intend to explore whether it   exhibits the  Van der Waals-like phase transition as the entanglement entropy.

We choose the quintessence  Reissner-Nordstr\"{o}m-AdS  black hole as the gravity background. Quintessence dark energy  model  is an  important model that can explain the acceleration expansion of  our universe. The black hole is a crucial component of the universe, so it will be interesting to exploring the effect of quintessence dark energy on the properties of black holes.
Some attempts about this topic have been done until now. In~\cite{Sb1,Sb2}, the quasinormal modes and Hawking radiation of the black holes surrounded by quintessence have been studied.   In~\cite{Libh}, the phase structure of the quintessence  Reissner-Nordstr\"{o}m-AdS  black hole has been investigated in the extended  phase space.
 Especially recently, there are also some works to study the quintessence AdS black hole in the framework of holography.   \cite{Chen:2012mva} discussed effect of the quintessence dark energy on the formation of the superconductor.  \cite{Zeng20151} studied the influence of quintessence dark energy  on the non-equilibrium thermalization.
  At present, although the dual field theoretical interpretation about the quintessence is still unclear, with the further investigations, one can get better understanding. With this motivation, we will study  the phase structure of the
   quintessence  Reissner-Nordstr\"{o}m-AdS  black hole

This paper is organized as follows. In the next section, we will discuss the thermal entropy phase transition in the space time dominated by the quintessence dark energy. We mainly concentrate on  phase transition in $T-S$ plane in a fixed charge ensemble, which is shown to be the Van der Waals-like phase transition. In Section~\ref{Nonlocal_observables}, we  study phase transition of the holographic entanglement entropy and two point correlation function respectively and find  that that both of them exhibit the Van der Waals-like phase transition.  Particularly for each  observable, the  equal area law is checked and the critical exponent of the heat capacity is obtained.
The last section is devoted to discussions and conclusions.

\section{Thermodynamics  of the quintessence  Reissner-Nordstr\"{o}m-AdS  black hole}
\subsection{Quintessence  Reissner-Nordstr\"{o}m-AdS  black hole}
\label{quintessence_Vaidya_AdS}
For a space time dominated by the quintessence dark energy with energy-momentum tensor
\begin{equation}
T_t^t=\rho_q(r),
\end{equation}
\begin{equation}
T_j^j=3\rho_q(r)\omega[-(1+3B)\frac{r^ir_j}{r_nr^n}+B\delta ^j_i],
\end{equation}
where  $\rho_q$ is dark energy density,  $\omega$ is state parameter, and $B$ is an arbitrary parameter
depending on the internal structure of quintessence, the charged AdS black hole solution can be written as \cite{Kiselev}
\begin{equation}
 ds^{2}=-f(r)dt^{2}+f^{-1}(r)dr^{2}+r^{2}(d\theta^2+\sin^2\theta d\phi^2),\label{metric1}
\end{equation} where%
\begin{equation}
 f(r)=1-\frac{a}{r^{3 \omega+1}}+\frac{r^2}{l^2}-\frac{2 M}{r}+\frac{Q^2}{r^2},\label{metric}
\end{equation}%
in which $l$ is the AdS radius which will be set to 1 during the numerics, $a$ is the normalization factor which  relates
to the density of quintessence with the relation
\begin{equation}
\rho_q =-\frac{a}{2}\frac{3\omega}{r^{3(\omega+1)}}.
\end{equation}
In cosmology, it is well known that for quintessence dark
energy $-1 <\omega < -1/3$, while for phantom dark energy $\omega< -1$.
 Mathematically, from ~Eq.(\ref{metric}), we know that for different values of state parameter $\omega$, the dark energy has different effect on the space time. For $\omega= -1$, the dark energy affects the AdS radius that is related to the cosmological constant. While for $\omega= -1/3$, the dark energy affects the curvature $k$ of the space time.

From ~Eq.(\ref{metric}), we can get the Hawking temperature of this space time
\begin{equation}
T_{bh}=\frac{f^{\prime}(r)}{4\pi}\mid_{r_+}=\frac{r_+ \left(3 a \omega r_+^{-3 \omega}+3 r_+^3/l^2+r_+\right)-Q^2}{4 \pi  r_+^3},\label{temperature}
 \end{equation}
which is regarded as the temperature of the dual conformal field theory according to AdS/CFT duality.  In addition,  according to the entropy area relation, we also can get the entropy of the black hole
\begin{equation}
S=\pi r_+^2,\label{entropy}
 \end{equation}
in above equations   $r_+$ is the event horizon of the black hole, which is the largest root of $f(r_+)=0$.

\subsection{Van der Waals-like phase transition of black hole entropy}
Substituting (\ref{entropy}) into (\ref{temperature}) and eliminating the parameter $r_+$, we can get the relation between the temperature $T_{bh}$ and entropy $S$ of the quintessence  Reissner-Nordstr\"{o}m-AdS  black hole, that is
\begin{equation}
T_{bh}=-\frac{S^{-\frac{3 \omega}{2}-\frac{3}{2}} \left(-3 a l^2 \sqrt{S} \pi ^{\frac{3 \omega}{2}+\frac{3}{2}} \omega+\pi ^2 l^2 Q^2 S^{\frac{3 \omega}{2}}-\pi  l^2 S^{\frac{3 \omega}{2}+1}-3 S^{\frac{3 \omega}{2}+2}\right)}{4 \pi ^{3/2} l^2}.\label{temperatures}
\end{equation}
Based on this relation, we will study the Van der Waals-like phase transition in the $T-S$ plane. In fact, it has been shown that there also exist Van der Waals-like phase transitions in the  $P-V$ plane, $T-S$ plane, and  $\Phi-Q$ plane for a charged black hole in the extended phase space, in which the cosmological constant is treated as a thermal variable.  To compare with the phase transition of entanglement entropy and some other nonlocal observables directly, we focus on only the phase transition of thermal entropy in the  $T-S$ plane in this paper. We will also pay attention to the affect of  both  $\omega$ and $a$ on the phase structure of the black hole. We take $\omega=-1, -2/3$ and $a=0.5/2, 1.1/2$ as examples.

In order to understand the phase transition in the  $T-S$ plane, we should first find the critical charge, which is  determined by the following equations
\begin{equation}
(\frac{\partial T_{bh}}{\partial S})_Q=(\frac{\partial^2 T_{bh}}{\partial S^2})_Q=0. \label{heat}
 \end{equation}
Inserting   (\ref{temperatures}) into (\ref{heat}),
we find for $\omega=-1$, the critical charge, critical entropy and critical temperature can be expressed as
\begin{equation}
Q_c=\frac{l}{6 \sqrt{1-a l^2}},
 \end{equation}
\begin{equation}
S_c=\frac{\pi  l^2}{6 \left(1-a l^2\right)}, \label{criticalentropy}
 \end{equation}
\begin{equation}
T_c=\frac{0.259899 a^2 l^4-0.519798 a l^2+0.259899}{\sqrt{\frac{l^2}{1-a l^2}} \left(1- a l^2\right)^2} \label{ct},
 \end{equation}
and for $\omega=-2/3$, these quantities are
\begin{equation}
Q_c= \frac{l}{6},
 \end{equation}
\begin{equation}
S_c=\frac{\pi  l^2}{6}, \label{1criticalentropy}
 \end{equation}
\begin{equation}
T_c=\frac{2 \sqrt{6}-3 al}{6 \pi
l }\label{1ct}.
 \end{equation}
All these critical values are useful for us to study the critical behavior of heat capacity near the critical point next.
\begin{figure}
\centering
\subfigure[$\omega=-2/3, a=0.5/2$]{
\includegraphics[scale=0.75]{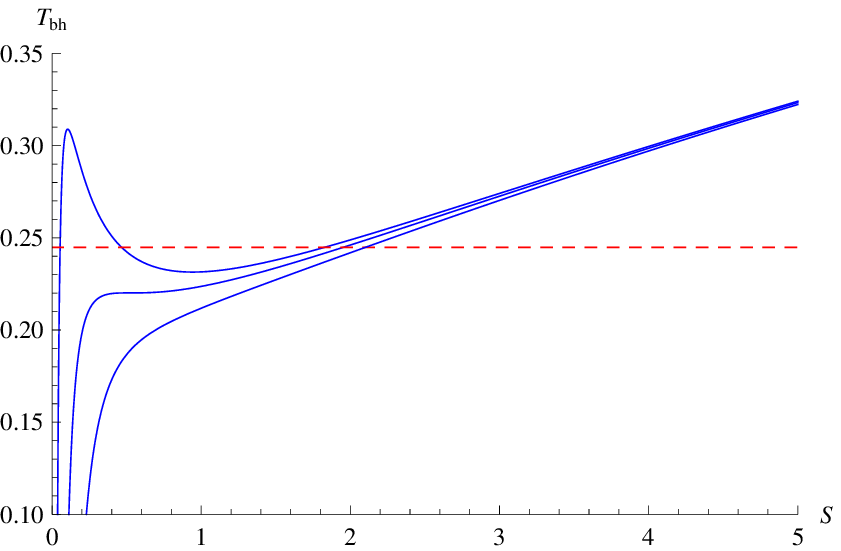}  }
\subfigure[$\omega=-2/3, a=1.1/2$]{
\includegraphics[scale=0.75]{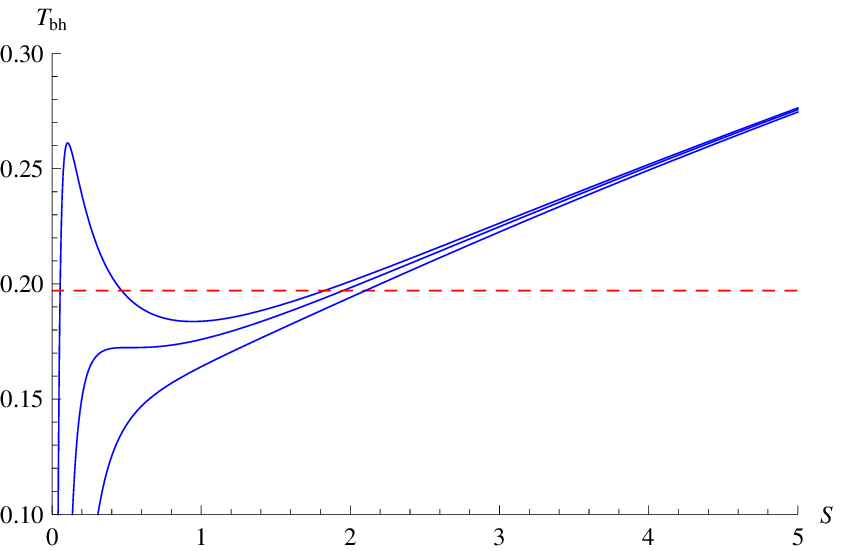}
}
\subfigure[$\omega=-1, a=0.5/2$]{
\includegraphics[scale=0.75]{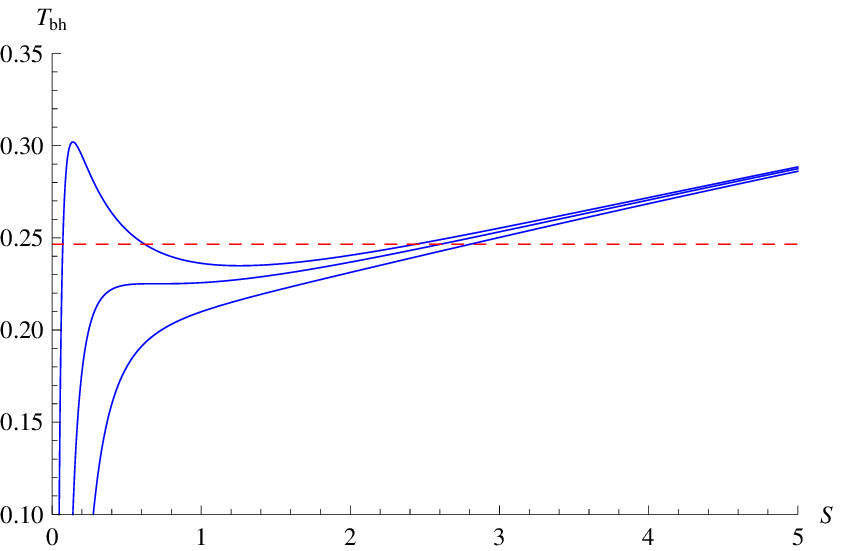}  }
\subfigure[$\omega=-1, a=1.1/2$]{
\includegraphics[scale=0.75]{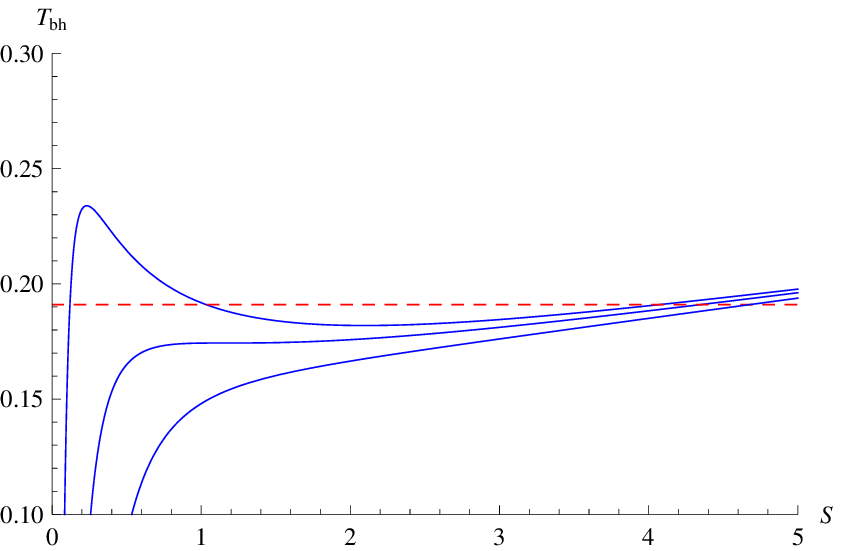}
}
 \caption{\small  Relations between the entropy and temperature  for different $\omega$ and $a$. The red dashed  lines  correspond to the locations of first order phase transition.} \label{fig1}
\end{figure}

With (\ref{temperatures}), we now plot the isocharges for different $\omega$ and $a$, which are shown in Figure \ref{fig1}. In (a) and (b) of  Figure \ref{fig1}, the curves from top to down correspond to the isocharges for the case $Q=0.6/6$, $1/6$, $1.4/6$. In (c) and (d),  these curves correspond to $Q=\frac{0.6}{6 \sqrt{1-a }}$, $\frac{1}{6 \sqrt{1-a }}$, $\frac{1.4}{6 \sqrt{1-a }}$.
It is obvious that for a fixed $\omega$ and $a$, the phase structure
of the quintessence  Reissner-Nordstr\"{o}m-AdS  black hole is
similar to that of the Van der Waals phase transition.
 That is, for the small charge, which corresponds to the top curves in each graph in Figure \ref{fig1}, there is an unstable hole interpolating between the stable small hole and stable large
 hole.  The small stable hole will jump to the large stable hole as the temperature of the hole is larger than the critical temperature $T_{\star}$.
  As the charge increases to the critical charge, the smallest hole and the largest hole merge into one and squeeze  out the unstable phase. So there is an inflection point in the middle curves for each graphic in Figure \ref{fig1}.
  According to the definition of the specfic heat capacity
  \begin{equation}
C_{Q}=T_{bh}\frac{\partial S}{\partial T_{bh}}\mid_Q \label{capacity},
 \end{equation}
we know that the heat capacity is divergent at the inflection point and the phase transition near there is  second order. As the charge exceeds the critical charge, the black hole is stable always, which correspond to  the lowest curves in each graph in Figure \ref{fig1}.
In addition, from Figure \ref{fig1}, we can also observe how  $\omega$ and $a$ affect the phase structure. From the top curves in (b) and (d), we know that as $\omega$ decreases, the unstable stage is longer. So the large $\omega$  promotes the hole to reach the stable stage.  From the top curves in (c) and (d), we know that the large $a$ delays the hole to reach the stable stage.

\begin{figure}
\centering
\subfigure[$\omega=-2/3, a=0.5/2,Q=(1-0.4)/6$]{
\includegraphics[scale=0.75]{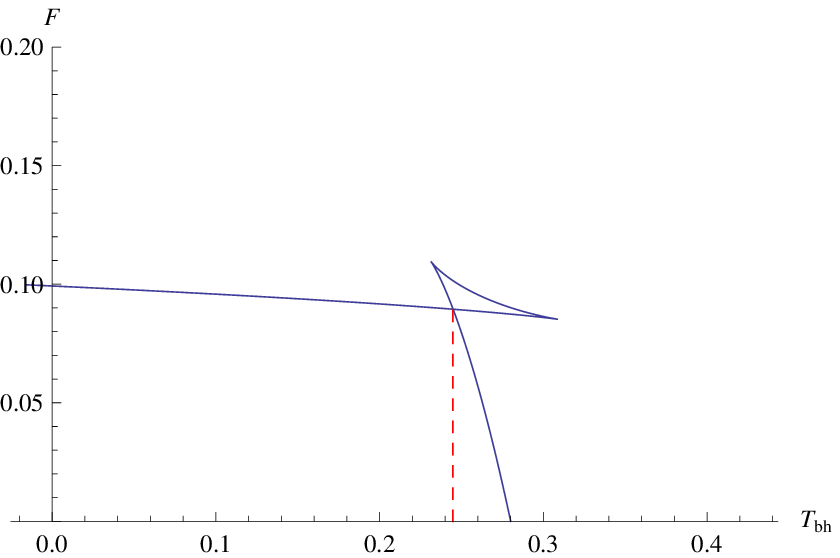}  }
\subfigure[$\omega=-2/3, a=1.1/2,Q=(1-0.4)/6$]{
\includegraphics[scale=0.75]{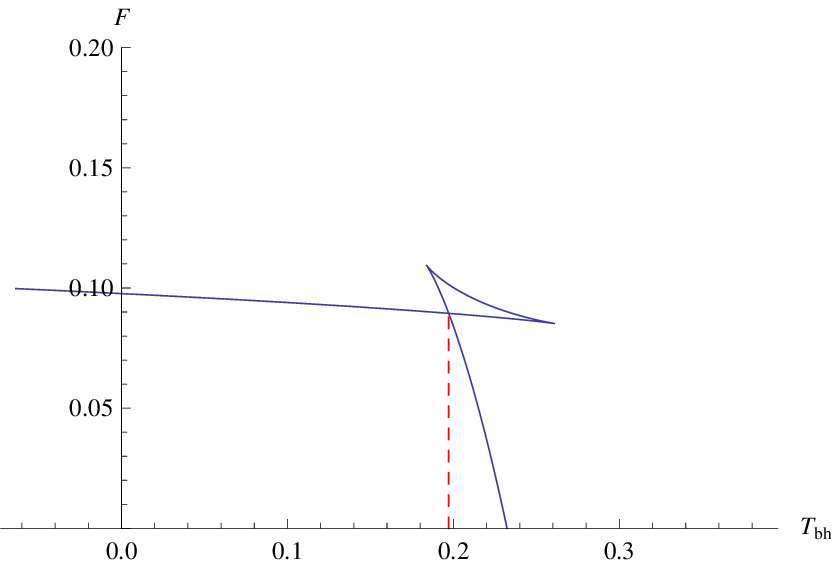}
}
\subfigure[$\omega=-1, a=0.5/2,Q=\frac{1-0.4}{6 \sqrt{1-a}}$]{
\includegraphics[scale=0.75]{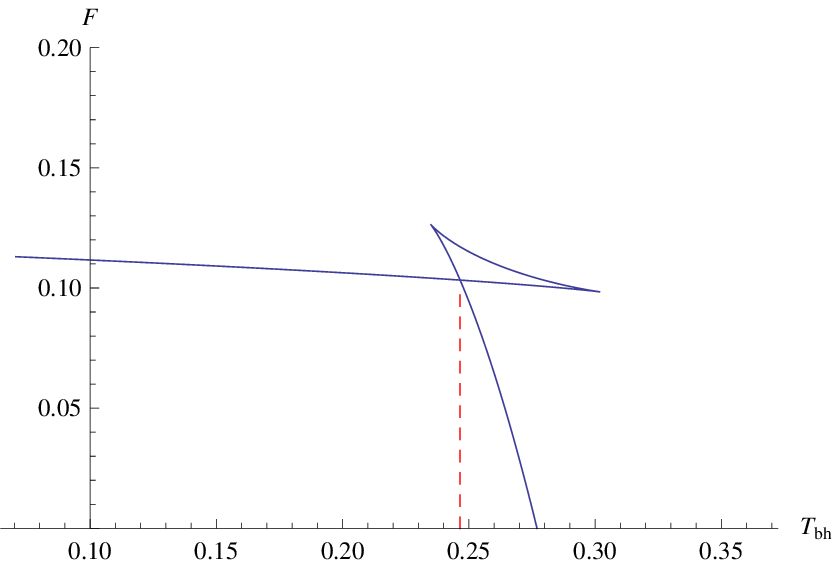}  }
\subfigure[$\omega=-1, a=1.1/2,Q=\frac{1-0.4}{6 \sqrt{1-a}}$]{
\includegraphics[scale=0.75]{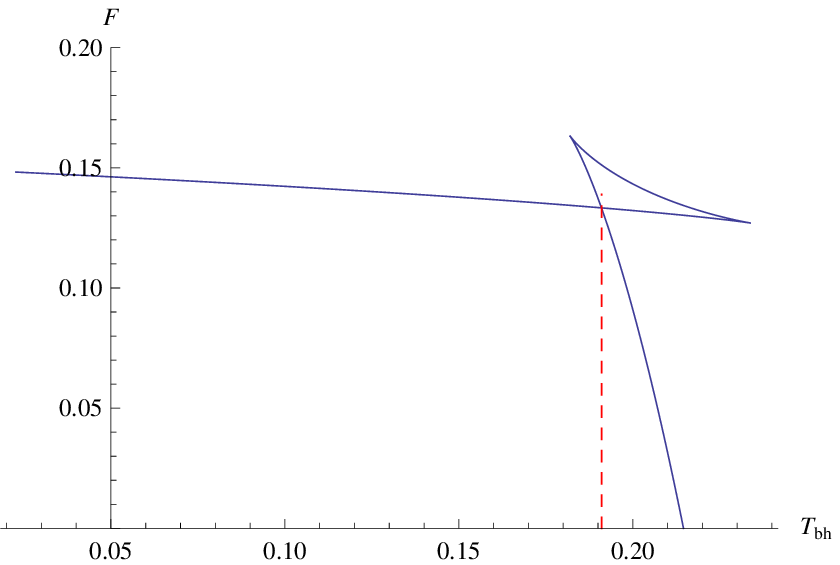}
}
 \caption{\small Relations between the free energy and temperature for different $\omega$ and $a$. The intersection point between the red dashed  line and horizontal coordinate  in each graph is the first order phase transition temperature.} \label{fig2}
\end{figure}

For the first order phase transition in Figure \ref{fig1}, we will check whether Maxwell's equal area law holds, which states
\begin{equation}
A_1\equiv\int_{S_1}^{S_3}T_{bh}(S,Q)dS=T_{\star}(S_3-S_1)\equiv A_3. \label{euqalarea}
 \end{equation}
Obviously, to check this equation, we should first find the value of the critical temperature $T_{\star}$. Usually there are two different ways to get it. On one hand, one can construct an equation that produces the temperature with the supposition that the equal area law is true. On the other hand, one can find the  horizontal coordinate of the junction of the swallowtail structure in the $F-T$ plane, where $F=M-TS$ is the Helmholtz free energy. Here our goal is to check the equal area law and the second method thus is more appropriate. The relations between $F$ and $T$ for different $\omega$ and $a$ are plotted in Figure \ref{fig2}. We can see that there is always a swallowtail structure in each graph, which corresponds to the unstable stage of the first order phase transition in Figure \ref{fig1}.
Now we take  $\omega=-2/3$, $a=0.5/2$ as an example to show how to check the equal area law. From (a) in Figure \ref{fig2}, we find  $T_{\star}=0.2448$.
Substituting this temperature into  (\ref{temperatures}), we get the smallest and largest values of the entropy, $S_1=0.210442$,  $S_3=2.93091$. With these values, we find $A_1$ and $A_3$ in (\ref{euqalarea}) equal  0.4340, and 0.4338 respectively.
Adopting  the similar strategy, we  can get  $A_1$ and $A_3$ for other  $\omega$ and $a$, which are shown in Table \ref{tab1}.
\begin{table}
\begin{center}\begin{tabular}{l|c|c|c|c|c}
 \hline
               &$w=-2/3$ &          $w=-1$         \\ \hline
 $a$=0.5/2   & $T_{\star}$=0.2448&   $T_{\star}$ =0.2465      \\ \hline
$a$=1.1/2   &$T_{\star}$=0.1971&   $T_{\star}$  =0.19103           \\ \hline
$a$=0.5/2   & $S_1$=0.053875$\mid$$ S_3$= 1.82594&   $S_1$=0.071847$\mid$ $ S_3$=2.43723       \\ \hline
$a$=1.1/2   &$S_1$=0.053886$\mid$ $S_3$=1.82799&   $S_1$=0.119813$\mid$  $S_3$=4.07542             \\ \hline
$a$=0.5/2   &$A_1$=0.4340$\mid$  $A_3$=0.4338&   $A_1$=0.5832$\mid$  $A_3$=0.5831      \\ \hline
$a$=1.1/2   &$A_1$=0.4398$\mid$  $A_3$=0.3497&   $A_1$=0.7555$\mid$  $A_3$=0.7557             \\ \hline

\end{tabular}
\end{center}
\caption{Check of the equal area law in the $T-S$ plane for different $\omega$ and $a$.}\label{tab1}
\end{table}
It is obvious that $A_1$ equals $A_3$ roughly for different $\omega$ and $a$, so the equal area law holds. In other words,
 though  $\omega$ and $a$ affect the phase structure of thermal entropy, it does not  break the equal area law.

For the second order phase transition, we are interested in the critical exponent associated with the heat capacity defined in
 (\ref{capacity}). We take $\omega=-2/3$ as an example. Near the critical point, writing the entropy as $S=S_c+\delta$  and expanding the temperature in small $\delta$, we find
\begin{eqnarray}
T_{bh}-T_c=\frac{ \left((35 \pi ^2 l^2 Q^2)/S_c-5 \pi  l^2+3 S_c\right)}{64 \pi ^{3/2} l^2 S_c^{7/2}}(S-S_c)^3\label{c1},
 \end{eqnarray}
in which  we have used (\ref{heat}).
With the definition of the heat capacity, we find further $C_Q\sim(T_{bh}-T_c)^{-2/3}$, namely the critical exponent is $-2/3$, which is the same as the one from the mean field theory. In addition taking  logarithm to (\ref{c1}), we find there is always a linear relation
\begin{eqnarray}
\log\mid T_{bh}-T_c\mid =3 \log\mid S- S_c\mid +constant  \label{c2},
 \end{eqnarray}
with 3 the slope.
Next, we will employ this relation to check the critical exponent of heat capacity  in entanglement entropy-temperature plane as well as two point correlation function-temperature plane.

\section{Van der Waals  Phase transition in the framework of holography}
\label{Nonlocal_observables}

Having obtained the phase structure of thermal entropy of the quintessence  Reissner-Nordstr\"{o}m-AdS  black hole, we will study the phase structure of the non-local observables such as entanglement entropy and two point  correlation function in the filed theory. We intend to explore whether they have the similar phase structure and critical behavior as that of the thermal entropy.

\subsection{Van der Waals  Phase transition of entanglement entropy}
 According to the formula in~\cite{Ryu,Ryu1}, the entanglement entropy can be given by the area
$A_{\Sigma}$ of a minimal surface $\Sigma$ anchored on $\partial \Sigma$, to wit
\begin{equation}
S=\frac{A_{\Sigma}(t)}{4 \pi G}  \label{eee},
\end{equation}
where $G$ is the  Newton's constant. Based on the definition of area and  (\ref{metric1}),
(\ref{eee}) can be rewritten as
\begin{eqnarray}
S=\frac{ \pi}{2} \int_0 ^{\theta_0}r \sin\theta\sqrt{\frac{(r^{\prime})^2}{f(r)}+r^2},
 \end{eqnarray}
in which $r^{\prime}=dr/ d\theta$  and $\theta_0$ is the boundary of the entangling region in $\theta$ direction.
 Making use of the Euler-Lagrange equation, one can get the equation of motion of $r(\theta)$
\begin{eqnarray}
0&=&r'(\theta )^2 [\sin \theta  r(\theta )^2 f'(r)-2 \cos \theta r'(\theta )]-2 r(\theta ) f(r) [r(\theta ) (\sin \theta  r''(\theta ) \nonumber \\
&+&\cos \theta  r'(\theta ))
-3 \sin \theta  r'(\theta )^2]+4 \sin (\theta ) r(\theta )^3 f(r)^2.
\end{eqnarray}
It seems to be impossible to get the analytical solution of  $r(\theta)$, so we will solve it numerically with the boundary conditions
\begin{eqnarray}
r^{\prime}(0)=0, r(0)= r_0 \label{bc}.
 \end{eqnarray}
 Note that the entanglement entropy is divergent at the boundary, so it should be regularized by subtracting off the entanglement entropy in pure AdS with the same entangling surface and boundary values.
We label the regularized entanglement entropy as $\delta S$.
For the numerical computation, we choose  $\theta_0=0.16$ and set the UV cutoff in the dual field theory  to be $r(0.159)$.
\begin{figure}
\centering
\subfigure[$\omega=-2/3, a=0.5/2$]{
\includegraphics[scale=0.75]{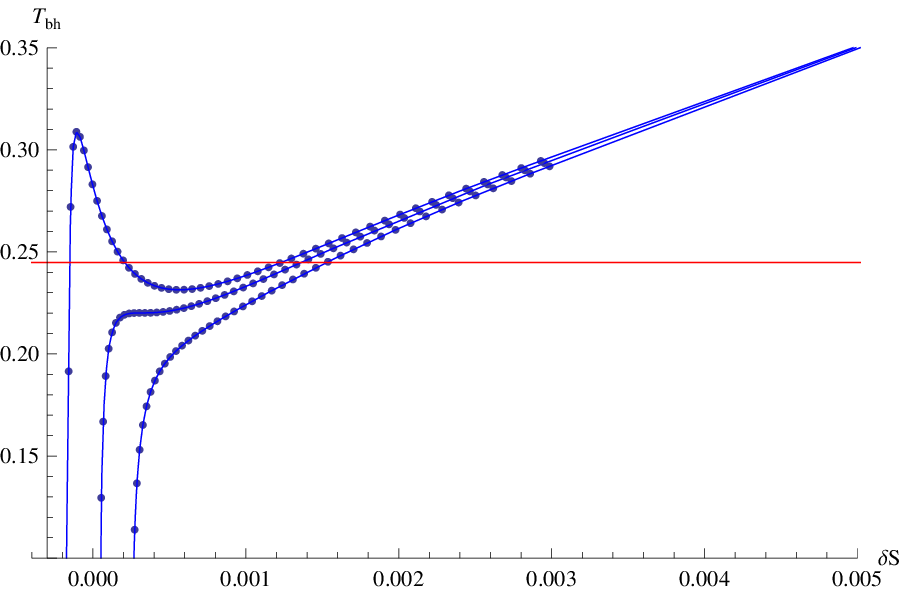}  }
\subfigure[$\omega=-2/3, a=1.1/2$]{
\includegraphics[scale=0.75]{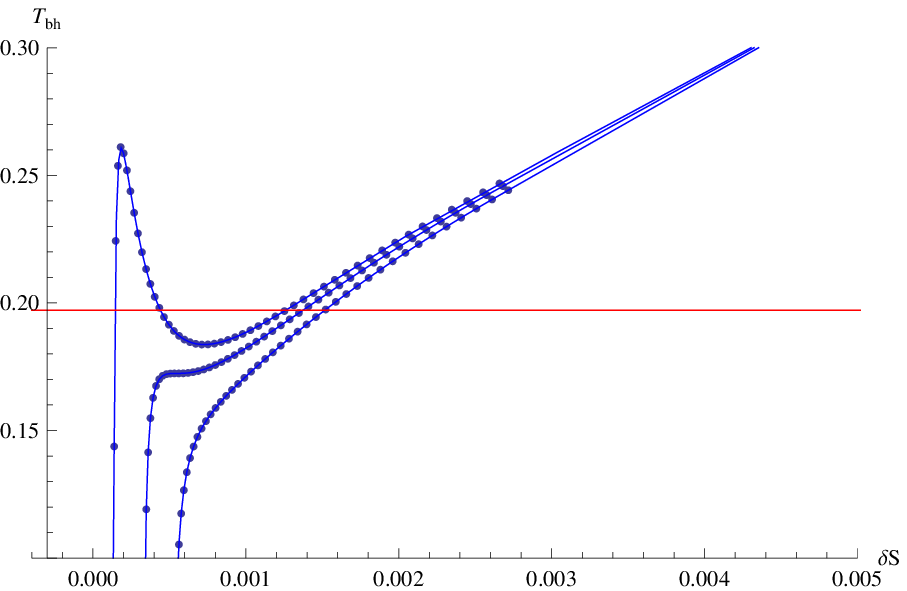}
}
\subfigure[$\omega=-1, a=0.5/2$]{
\includegraphics[scale=0.75]{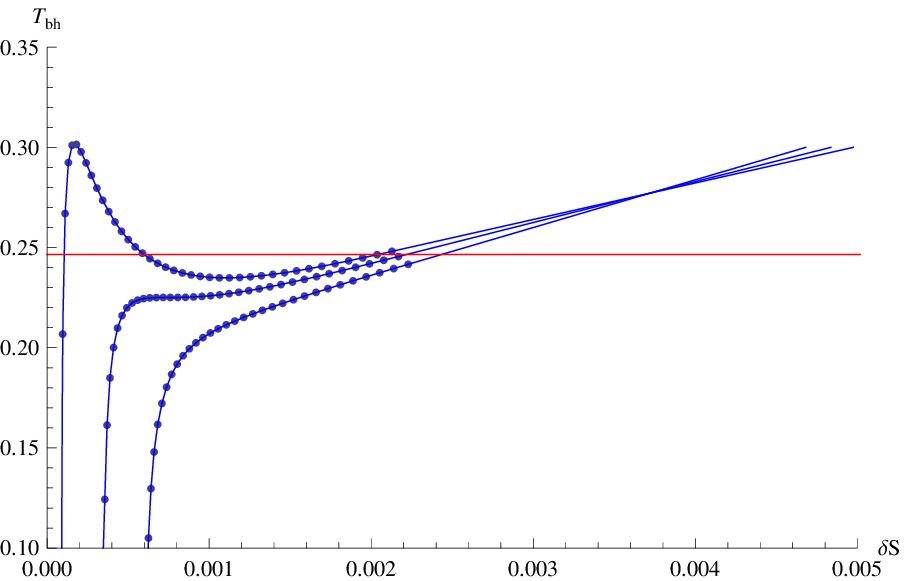}  }
\subfigure[$\omega=-1, a=1.1/2$]{
\includegraphics[scale=0.75]{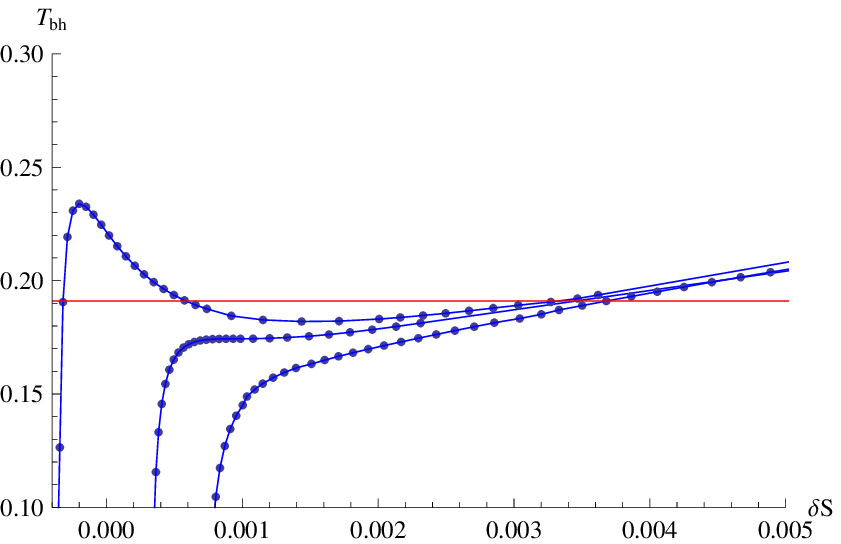}
}
 \caption{\small Relations between the entanglement entropy and temperature   for different $\omega$ and $a$. The red solid lines  correspond to the locations of first order phase transition.} \label{fig3}
\end{figure}
To compare with the  phase transition of thermal entropy, we will study the relation between the entanglement entropy and Hawking temperature, which is regarded as the temperature of the dual field theory. The numeric results for different  $\omega$ and $a$ are shown in Figure \ref{fig3}. As the same as that of the thermal entropy, in (a) and (b) in Figure \ref{fig3}, the curves from top to down correspond to the isocharges for the case $Q=0.6/6$, $1/6$, $1.4/6$ individually. In (c) and (d),  these curves correspond to $Q=\frac{0.6}{6 \sqrt{1-a }}$, $\frac{1}{6 \sqrt{1-a }}$, $\frac{1.4}{6 \sqrt{1-a }}$.
For a fixed  $\omega$ and $a$, we find the phase structure of entanglement entropy is the same as that of the thermal entropy. Namely the phase transition is similar to that of the Van der Waals  Phase transition.
In other words, the phase structure depends on the charge of the black hole. For the small charge, the small stable black hole will transfer to the large stable hole as the temperature is higher than the critical temperature, and this transition is first order. As the charge grows to the critical charge, the small hole and the large hole merge so that the unstable hole shrinks into an inflection point, where the  transition for the small hole to the large hole is second order. For a large enough charge, a large stable hole forms and the entanglement entropy grows monotonously as the temperature rises. From Figure \ref{fig3}, we can also observe how  $\omega$ and $a$ affect the phase structure of entanglement entropy. From the top curves in (b) and (d), we know that as $\omega$ decreases, the unstable stage is longer, and from the top curves in (c) and (d), we know that  as $a$ decreases, the unstable stage is shorter. These effects are the same as that of the black hole entropy in  Figure \ref{fig1}.

Adopting the same strategy as that of the thermal entropy, we will check  Maxwell's equal area law for the first order phase transition.
In the  $\delta S-T$ plane, we rewrite the  equal area law as
\begin{eqnarray}
A_1\equiv\int_{\delta S_1}^{\delta S_3} T(\delta S) d\delta S= T_{\star} (\delta S_3-\delta S_1)\equiv A_3,\label{arealaw}
 \end{eqnarray}
in which  $T(\delta S)$  is an Interpolating Function obtained from the numeric result, and $\delta S_1$,  $\delta S_3$ are the smallest and largest roots of the equation $T(\delta S)=T_{\star}$. Surely $ T_{\star} $ is the first order phase transition temperature for a fixed  $\omega$ and $a$, which can be read off from  Figure \ref{fig2}.
For different  $\omega$ and $a$, the results of  $\delta S_1$,  $\delta S_3$  and $A_1$,  $A_3$ are listed in Table
(\ref{tab2}).
\begin{table}
\begin{center}\begin{tabular}{l|c|c|c|c|c}
 \hline
               &$w=-2/3$ &          $w=-1$         \\ \hline
 $a$=0.5/2   & $T_{\star}$=0.2448&   $T_{\star}$ =0.2465
    \\ \hline
$a$=1.1/2   &$T_{\star}$=0.1971&   $T_{\star}$  =0.19103
     \\ \hline
$a$=0.5/2   & $\delta S_1$=0.00021$\mid$$\delta S_3$= 0.00123&   $\delta S_1$=0.000110$\mid$ $ \delta S_3$=0.002128  \\\hline
$a$=1.1/2   &$\delta S_1$=0.0001453$\mid$  $\delta S_3$=0.0012605&   $\delta S_1$=0.000731$\mid$  $\delta S_3$=0.003087            \\\hline
$a$=0.5/2   &$A_1$=0.000240$\mid$  $A_3$=0.000249&   $A_1$=0.000479$\mid$  $A_3$=0.000477
  \\ \hline
$a$=1.1/2   &$A_1$=0.000221$\mid$  $A_3$=0.000220&   $A_1$=0.000436$\mid$  $A_3$=0.000447
    \\ \hline
\end{tabular}
\end{center}
\caption{Check of the equal area law in the $T-\delta S$ plane for different $\omega$ and $a$.}\label{tab2}
\end{table}
It is obvious that $A_1$ equals nearly   $A_3$ for  a fixed $\omega$ and $a$. That is, the equal area law is also valid for the first order phase transition of entanglement entropy. This result is the same as that of the thermal entropy of the black hole.

\begin{figure}
\centering
\subfigure[$\omega=-2/3, a=0.5/2,Q=1/6 $]{
\includegraphics[scale=0.75]{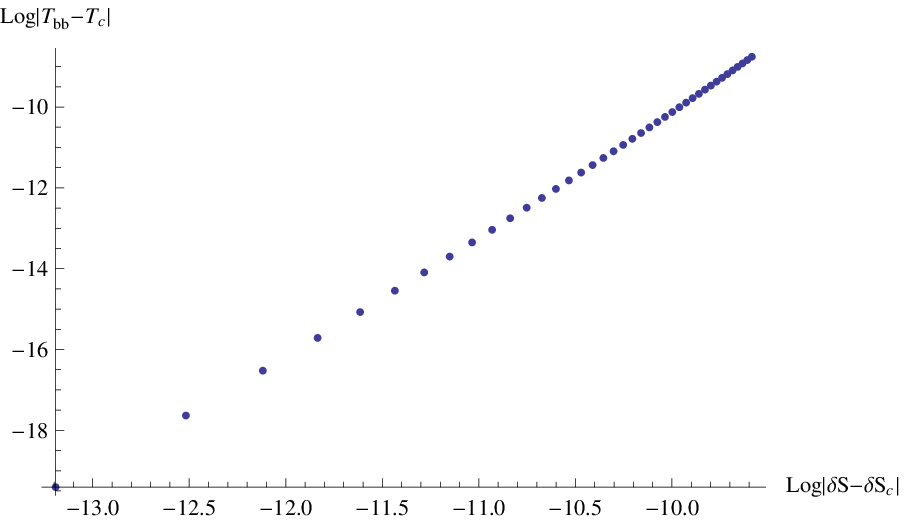}  }
\subfigure[$\omega=-2/3, a=1.1/2,Q=1/6 $]{
\includegraphics[scale=0.75]{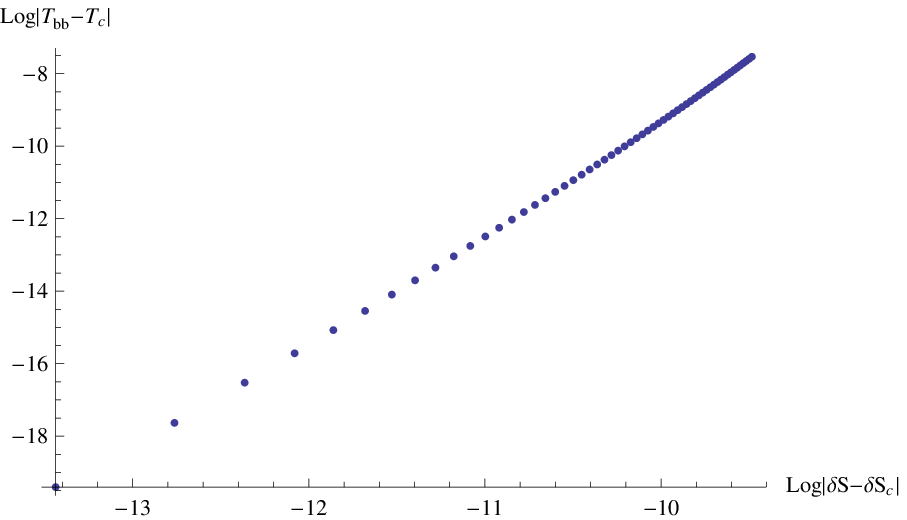}  }
\subfigure[$\omega=-1, a=0.5/2,Q=\frac{1}{6 \sqrt{1-a}}$]{
\includegraphics[scale=0.75]{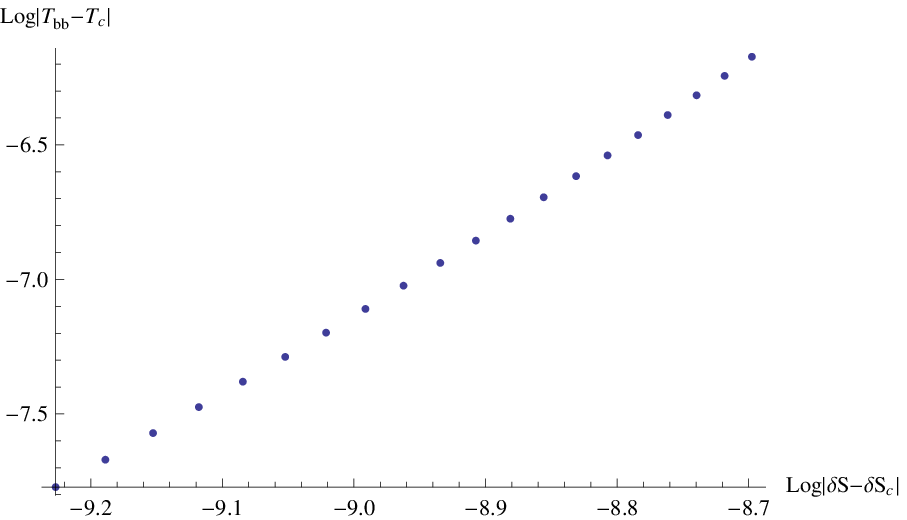}  }
\subfigure[$\omega=-1, a=1.1/2,Q=\frac{1}{6 \sqrt{1-a}}$]{
\includegraphics[scale=0.75]{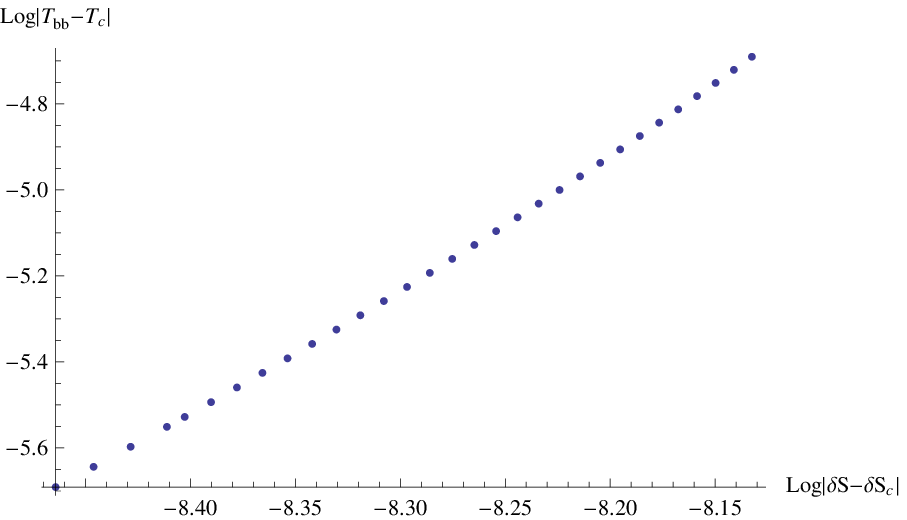}  }
 \caption{\small  Relations between  $\log\mid T_{bh}-T_c\mid$ and $\log\mid\delta S-\delta S_c\mid $ for different  $\omega$ and $a$. } \label{fig4}
\end{figure}

In order to study the critical exponent of the second order phase transition in the $\delta S-T$ plane, we define an analogous specific  heat capacity
\begin{eqnarray}
\mathcal{C}_Q=T_{bh}\frac{\partial \delta S}{\partial T_{bh}}\mid_Q. \label{cheat1}
 \end{eqnarray}
Provided a similar relation as that in (\ref{c2}) is satisfied, then with (\ref{cheat1}) we can get the critical exponent of second order phase transition of entanglement entropy.
Here  we are interested in the logarithm of the quantities $T_{bh}-T_c$, $\delta S-\delta S_c$.
For different  $\omega$ and $a$, the relation between $ \log\mid T_{bh}-T_c\mid$ and $\log\mid\delta S-\delta S_c\mid  $ are plotted in  Figure \ref{fig4},
in which $T_c$ is the critical temperature which have been defined in (\ref{ct}) as well as  (\ref{1ct}), and
 $S_c$ is the corresponding critical entropy obtained numerically by the equation $T(\delta S)=T_c$. The analytical result of these straight lines can be fitted as
\begin{equation}
\log\mid T_{bh}-T_c\mid=\begin{cases}
19.9074 + 3.00361 \log\mid\delta S-\delta S_c\mid, &$for$ ~\omega=-2/3,a=0.5/2 \\
21.5846 + 3.08801 \log\mid\delta S-\delta S_c\mid,&$for$ ~\omega=-2/3,a=1.1/2 \\
20.1355 + 3.02826  \log\mid\delta S-\delta S_c\mid,& $for$ ~\omega=-1,a=0.5/2\\
19.8945 + 3.02606 \log\mid\delta S-\delta S_c\mid. & $for$ ~\omega=-1,a=1.1/2\\
\end{cases}
\end{equation}
We find for all the $\omega$ and $a$, the slope is always about 3, which is consistent with that  of the thermal entropy.  That is, the entanglement entropy has the same second order phase transition behavior  as that of the thermal entropy.

\subsection{Van der Waals  Phase transition of two point correlation function}

\begin{figure}
\centering
\subfigure[$\omega=-2/3, a=0.5/2,Q=1/6$]{
\includegraphics[scale=0.75]{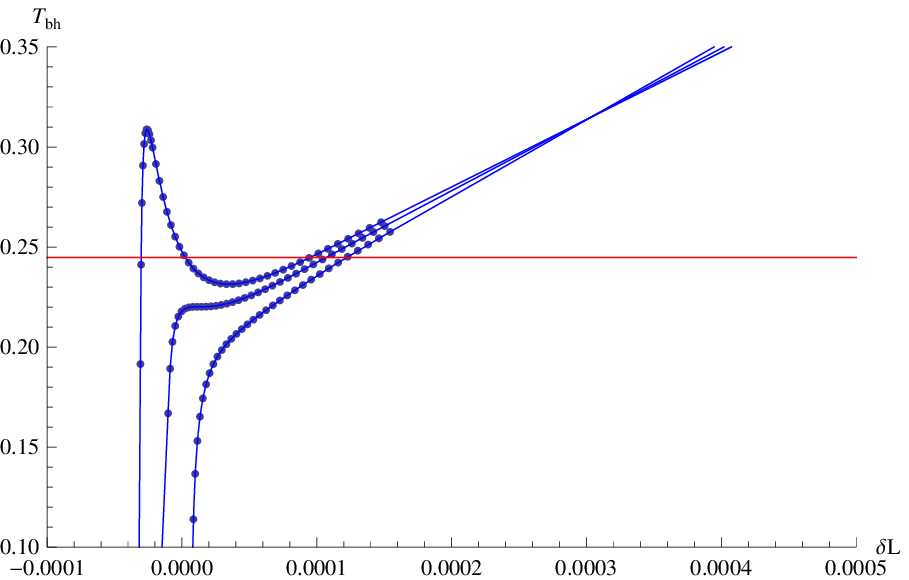}  }
\subfigure[$\omega=-2/3, a=1.1/2,Q=1/6$]{
\includegraphics[scale=0.75]{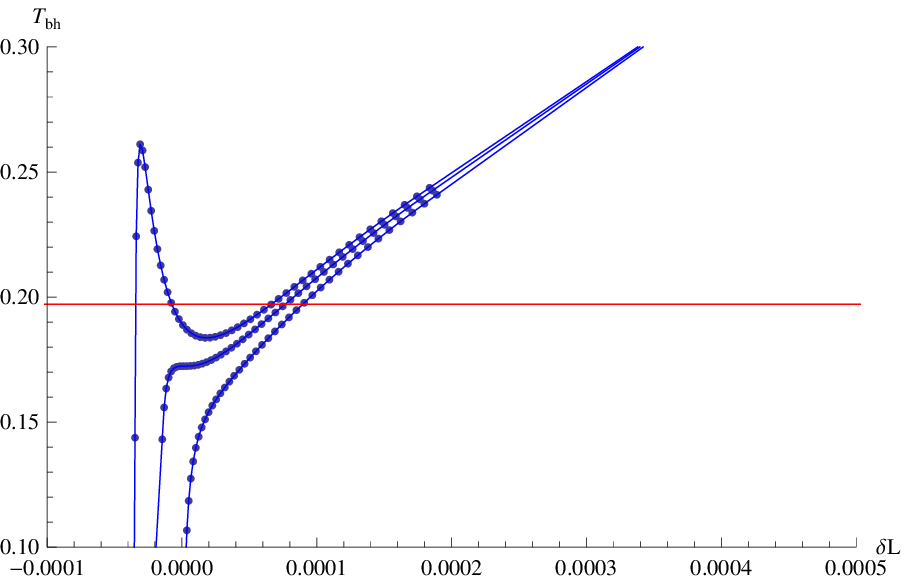}
}
\caption{\small Relations between the two point correlation function and temperature  for different $a$ with a   fixed  $\omega=-2/3$. The red solid  lines  correspond to the locations of first order phase transition.} \label{fig5}
\end{figure}

In previous section, we have shown that the entanglement entropy has the same phase structure as that of the thermal entropy. In this section, we intend to explore whether the two point correlation function has the similar behavior as that of the entanglement entropy. According to the AdS/CFT correspondence,  the equal time two point correlation function under the  the saddle-point
approximation can
be holographically approximated as \cite{Balasubramanian61}
\begin{equation}
\langle {\cal{O}} (t_0,x_i) {\cal{O}}(t_0, x_j)\rangle  \approx
e^{-\Delta {L}} ,\label{llll}
\end{equation}
if the conformal dimension $\Delta$ of scalar operator $\cal{O}$
is large enough, where
$L$ is the  length of the bulk geodesic between the points $(t_0,
x_i)$ and $(t_0, x_j)$ on the AdS boundary.
 Taking into account the spacetime symmetry of the quintessence  Reissner-Nordstr\"{o}m-AdS  black hole,
  we can simply let $x_i=\theta$ with the boundary $\theta_0$. We will also  employ  $\theta$  to
    parameterize the trajectory and in this case the proper length  is given by
\begin{eqnarray}
L=\int_0 ^{\theta_0}\mathcal{L}(r(\theta),\theta) d\theta,~~\mathcal{L}=\sqrt{\frac{(r^{\prime})^2}{f(r)}+r^2},
 \end{eqnarray}
in which $r^{\prime}=dr/ d\theta$.
Imagining $\theta$ as time, and treating $\mathcal{L}$ as the Lagrangian, one can
get the equation of motion for $r(\theta)$ by making use of the Euler-Lagrange equation, that is
\begin{eqnarray}
0=r'(\theta )^2 f'(r)-2 f(r) r''(\theta )+2 r(\theta ) f(r(\theta ))^2.
\end{eqnarray}
We will also use  ~Eq.(\ref{bc}) to solve this equation. We choose  $\theta_0=0.16$ and set the UV cutoff in the dual field theory  to be $ r(0.159)$.
We label the regularized two point correlation function as $\delta L$. For simplicity in this section, we fix $\omega=-2/3$. The relations between  $\delta L$ and $T_{bh}$  for different  $a$ ares shown in Figure \ref{fig5}.
 In each figure, the curves from top to down correspond to  the isocharges for $Q=0.6/6$, $1/6$, $1.4/6$. Comparing Figure \ref{fig5} with (a) and (b) in Figure \ref{fig3}, we find they are the same nearly. That is, the phase structure of the two point correlation function is also similar to that of the Van der Waals-like phase transition. To further confirm this conclusion, we will also pay attention to the equal area law for the first order phase transition and critical exponent for the second order phase transition.
  In the  $\delta L-T$ plane, we define the  equal area law  as
\begin{eqnarray}
A_1\equiv\int_{\delta L_1}^{\delta L_3} T(\delta L) d\delta L= T_{\star} (\delta L_3-\delta L_1)\equiv A_3,\label{arealaw}
 \end{eqnarray}
in which  $T(\delta L)$  is an Interpolating Function obtained from the numeric result, and $\delta L_1$,  $\delta L_3$ are the smallest and largest roots of the equation $T(\delta L)=T_{\star}$.
\begin{table}
\begin{center}\begin{tabular}{l|c|c|c|c|c}
 \hline
               &$w=-2/3$                 \\ \hline
 $a$=0.5/2   & $T_{\star}$=0.2448       \\ \hline
$a$=1.1/2   &$T_{\star}$=0.1971             \\ \hline
$a$=0.5/2   & $\delta L_1$=0.000002483098 $\mid$ $ \delta L_3$=0.000095059       \\ \hline
$a$=1.1/2   &$\delta L_1$=-0.000007475310 $\mid$  $\delta L_3$=0.000066343            \\\hline
$a$=0.5/2   &$A_1$=0.0000220$\mid$  $A_3$=0.0000227  \\ \hline
$a$=1.1/2   &$A_1$=0.0000140$\mid$  $A_3$=0.0000146             \\ \hline

\end{tabular}
\end{center}
\caption{Check of the equal area law in the $T-\delta L$ plane for different $a$ .}\label{tab3}
\end{table}
For different  $a$, the results of  $\delta L_1$,  $\delta L_3$  and $A_1$,  $A_3$ are listed in Table
(\ref{tab3}).  From this table, we can see that $A_1$ equals   $A_3$ in our numeric accuracy.
It is obvious that the equal area law is also true in the $\delta L-T$ plane. In other words, similar to the entanglement entropy, the two point correlation function also exhibits the  first order phase transition as that of the thermal entropy.

\begin{figure}
\centering
\subfigure[$\omega=-2/3, a=0.5/2,Q=1/6$]{
\includegraphics[scale=0.75]{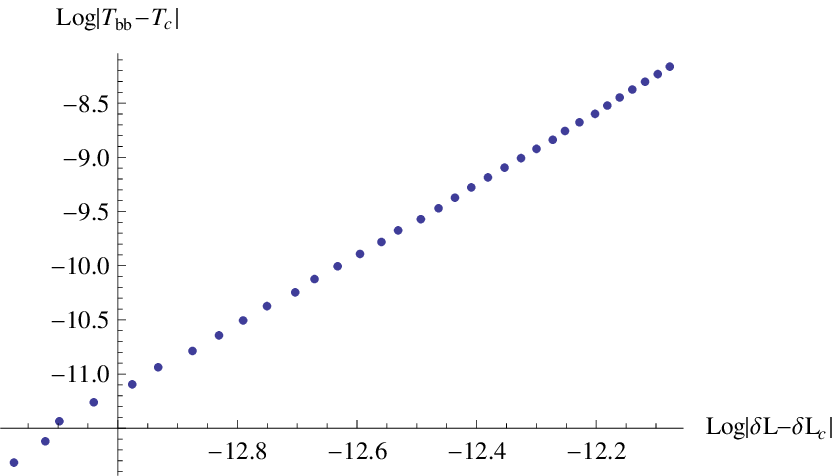}  }
\subfigure[$\omega=-2/3, a=1.1/2,Q=1/6$]{
\includegraphics[scale=0.75]{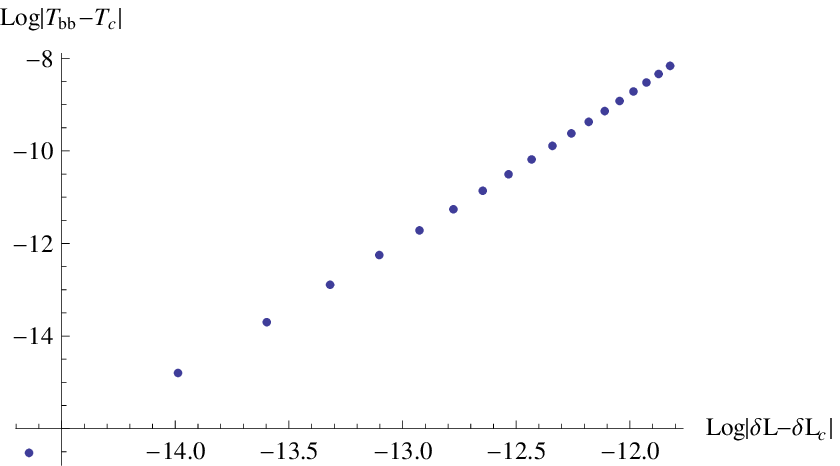}  }
\caption{\small  Relations between $\log\mid T_{bh}-T_c\mid$ and $\log\mid\delta L-\delta L_c\mid $ for different $a$ .} \label{fig6}
\end{figure}

By defining an
analogous   heat capacity
\begin{eqnarray}
C=T_{bh}\frac{\partial \delta L}{\partial T_{bh}}, \label{cheat2}
 \end{eqnarray}
we can also study the critical exponent of the heat capacity for the second order phase transition in the $\delta L-T$ plane.
Similar to that of the entanglement entropy,  we are interested in the logarithm of the quantities $T_{bh}-T_c$, $\delta L-\delta L_c$, in which $T_c$ is the critical temperature defined in  (\ref{1ct}), and
 $L_c$ is  obtained numerically by the equation $T(\delta L)=T_c$. The relations between $ \log\mid T_{bh}-T_c\mid$ and $\log\mid\delta L-\delta L_c\mid  $ are plotted in Figure \ref{fig6},
The analytical results of these
curves can be fitted as
\begin{equation}
\log\mid T_{bh}-T_c\mid=\begin{cases}
27.2601+3.00747 \log\mid\delta L-\delta L_c\mid,&  $for$ ~\omega=-2/3,a=0.5/2\\
31.0971+3.15294 \log\mid\delta L-\delta L_c\mid. &~$for$\omega=-2/3,a=1.1/2\\
\end{cases}
\end{equation}
It is obvious that the slope is also about 3, which is consistent with that  of the thermal entropy.  That is, the two point correlation function also has the same second order phase transition   as that of the thermal entropy.

\section{Concluding remarks}

We have investigated  detailedly the Van der Waals-like phase transition in the quintessence  Reissner-Nordstr\"{o}m-AdS  black hole in a fixed charge ensemble. We first investigated the phase structure of the thermal entropy in the $T-S$ plane and  found that the phase structure depends on the charge of the black hole. For the small charge, there is always an unstable hole interpolating between the small stable black hole and large stable black hole. The transition for the small hole to the large hole is first order and the equal area law holds. As the charge of the hole increases to the critical value, the unstable hole merges into an inflection point where the phase transition is second order. While the charge is larger than the critical charge, the hole is stable always. We also discussed the effect of $\omega$ as well as $a$ on the phase structure and found $\omega$  promote the hole to reach the stable phase while  $a$ delays.
Interestingly, we found that the nonlocal observables such as entanglement entropy and two point correlation function also exhibit the similar phase structure in the $T-\delta S$ plane and $T-\delta L$ plane respectively. The influence for
 $\omega$ and $a$  on the phase structure is the same as that of the thermal entropy. To confirm this observation, we further showed that the equal area law holds  and the critical exponent of the heat capacity is consistent with that of the mean field theory  for both  the entanglement entropy and two point correlation function. These results imply that the entanglement entropy and the two point correlation function are indeed a good probe to the phase transition.


\section*{Acknowledgements}
We are grateful  to Professor Rong-Gen Cai for his instructive discussions.
This work  is supported  by the National
 Natural Science Foundation of China (Grant No. 11405016, Grant No. 11205226, Grant No. 11575270), and  Natural Science
 Foundation of  Education Committee of Chongqing (Grant No. KJ1500530).

\end{document}